\documentclass[final,3p,times,twocolumn]{elsarticle}
\usepackage{graphicx}
\usepackage{epsfig}
\usepackage{amssymb}
\usepackage{amsmath}
\usepackage{color}

\journal{Physica C}

\begin{document}
\begin{frontmatter}

\title{Design and implementation of a micro-coil induction magnetometer}
\author[1]{R. V. Iba\~nez Bustos}
\author[1,2]{S. D. Calder\'on Rivero}
\author[1,2]{F. Rom\'a}
\author[3]{H. Pastoriza}
\author[1,2]{M. I. Dolz}
\address[1]{Department of Physics, Universidad Nacional de San Luis, D5700BWS San Luis, Argentina}
\address[2]{INFAP, CONICET, D5700BWS San Luis, Argentina}
\address[3]{Low Temperatures Division, Centro At\'omico Bariloche, CNEA,
CONICET, R8402AGP San Carlos de Bariloche, R\'{\i}o Negro, Argentina}

\begin{abstract}
We present a micron-sized induction magnetometer designed 
to measure the magnetic response of superconducting mesoscopic samples. 
The device is manufactured using the Memscap PolyMUMPs process 
and consists of two octagonal planar parallel micro-coils  
covering an area of $240 \ \mu$m $\times$ $240 \ \mu$m, which are separated by only $2.75 \ \mu$m.
We show that this design is sufficiently sensitive to detect 
the Meissner transition at zero dc field, of a high-$T_\textrm{c}$ 
superconductor Bi$_2$Sr$_2$CaCu$_2$O$_8$ disk  
of $40 \ \mu$m in diameter and $1 \ \mu$m in thicknesses.
\end{abstract}

\end{frontmatter}

%\date{\today}
%.......................................................................................
\section{Introduction}

The experimental study of microscopic magnetic systems is today possible due to 
the miniaturization of different types of sensors \cite{Korvink2006}.
At the mesoscopic level, these devices allow to determine how depends 
on the sample size a given observable (magnetization, susceptibility, etc), 
or enable to compare the static and the dynamic behavior of 
a nanostructured object (nanowires, nanotubes, etc) with that of the corresponding bulk material,
providing information about the main physical mechanisms that dominate on such small scales 
\cite{Zijlstra1970,Wernsdorfet1996,Todorovic1998,Dolz2007,Dolz2008,Dolz2010,Dolz2015}.   

Magnetic sensors are based on a variety of physical phenomena: 
electromagnetic induction, magneto-resistivity, Hall effect, 
Josephson effect (SQUID sensors), magneto-optic effect, and
mechanical resonance (micro-electro-mechanical sensors), among others \cite{Ripka2001,Lyshevski2002,Lenz2006,Tumanski2013}.   
In particular, induction sensors are relatively simple to design and build, 
and allow for the non-invasive measurement of the complex susceptibility 
(as compared with other sensors types which can disturb the measurement 
of the magnetic field) \cite{Tumanski2007}. 

Previously, a variety of induction sensors has been developed 
to study small magnetic samples \cite{Brien2002,Dolz2012}. 
In those designs, the coils have linear dimensions of order of millimeter
allowing to analyze samples about this size.
In this work, we study the performance of a micro-coil sensor with which it is possible 
to measure the magnetic response of micron-sized samples.
The device is manufactured using the Memscap PolyMUMPs process and, to test it, 
we use a disk of $40 \ \mu$m in diameter and $1 \ \mu$m in thicknesses,
of the high-$T_\textrm{c}$ superconductor Bi$_2$Sr$_2$CaCu$_2$O$_8$ (BSCCO).  

The paper is structured as follows. The design and manufacturing process  
of the device are described in Section \ref{design-fabrication},
while the experimental setup is presented in Section \ref{setup}. 
The main results are given in the Section \ref{results}.  
Finally, conclusions are drawn in Section \ref{conclusions}.    
   
%........................................................................................
\section{Micro-coil sensor \label{design-fabrication}}

\begin{figure}[t!]
\begin{center}
\includegraphics[width=8cm,clip=true]{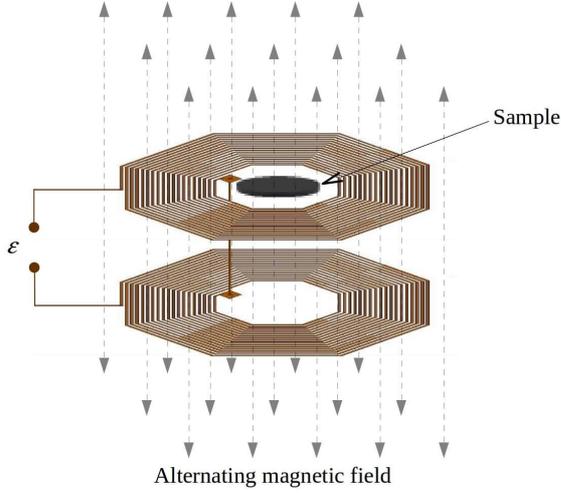}
\caption{Schematic diagram of the micro-coil sensor. }
\label{figure1}
\end{center}
\end{figure}

A schematic diagram of our micro-coil sensor is shown in Fig.~\ref{figure1}.  
Its working principle is very simple and is based on the Faraday's induction law.  
The two planar micro-coils have octagonal geometry and are concentric and parallel to each other.
They are connected in series at their inner ends  
and are subject to an alternating uniform magnetic field normal to them. 
Assuming that both are perfectly identical in size and shape, 
the total induced electromotive force, $\varepsilon$, will be zero.  
However, if we place a magnetic sample on top 
of the micro-coils (see Fig.~\ref{figure1}),
the system becomes unbalanced producing a voltage signal that can be measured.
  
To manufacture the device we use the multiuser commercial 
process PolyMUMPs of Memscaps Inc. \cite{Memscap}.  
This process offers seven sequential layers deposited on a silicon wafer: 
the bottom one is an isolation layer of silicon nitride; 
three polysilicon layers, Poly0, Poly1, and Poly2 with 
typical values of sheet resistance, 30, 10, and 20 $\Omega/$sq, respectively; 
and between them two sacrificial layers 
of phosphosilicate glass, Ox1 and Ox2 (insulating); 
on top of that one Metal layer of gold is provided.
All layers but the Silicon Nitride one are fully patternable by optical lithography and Reactive Ion etching, allowing the design of complex interconnected 3D structures.  
Table~\ref{TableMemscap} summarize the different layers and their thicknesses.           

\begin{table}[b!]
\begin{center}
\begin{tabular}{|l|c|}
\hline
Material layer        & Thicknesses ($\mu$m)       \\
\hline
Silicon nitride       & 0.6       \\
Polysilicon (Poly0)   & 0.5       \\
Phosphosilicate (Ox1) & 2.0       \\
Polysilicon (Poly1)   & 2.0       \\
Phosphosilicate (Ox2) & 0.75      \\
Polysilicon (Poly2)   & 1.5       \\
Gold (Metal)          & 1.5       \\
\hline
\end{tabular}
\caption{Layer names and thicknesses for the PolyMUMPs process.}
\label{TableMemscap}
\end{center}
\end{table}

\begin{figure}[t!]
\begin{center}
\includegraphics[width=8cm,clip=true]{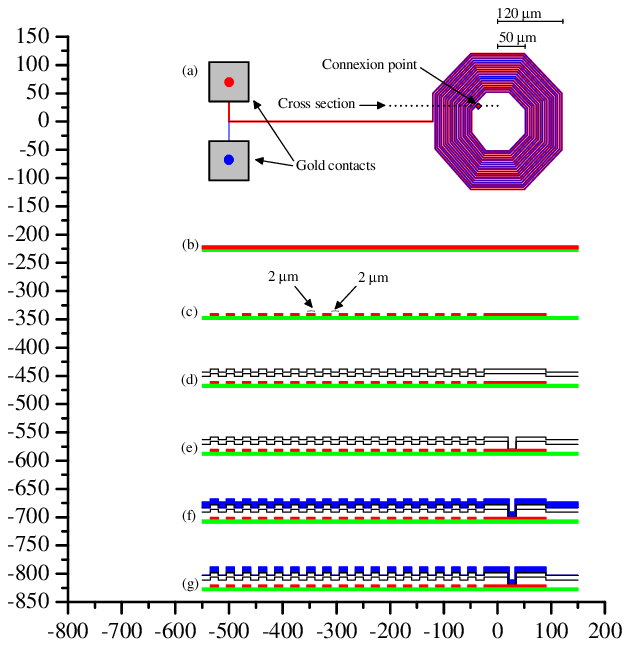}
\caption{(a) Schematic top view of the micro-sensor.  
Panel (b-g) show the main steps in the manufacturing process.  See text for details. }
\label{figure2}
\end{center}
\end{figure}

Figure~\ref{figure2} (a) presents a schematic top view of the device.
The micro-coils were built on Poly0 (red) and Poly2 (blue) layers.
For each one of them, we choose an octagonal geometry with $N=18$ turns whose apothems 
(the line drawn from the center of the octagon that is perpendicular to one of its sides)
ranged from $50$ to $120 \ \mu$m.  Wires have $2 \ \mu$m in width with a separation of $2 \ \mu$m between them.
The total resistance of the micro-sensor at room temperature is approximately $0.25$ M$\Omega$.  
  
The steps of the manufacturing process are shown in Figs.~\ref{figure2} (b-g),
and correspond to the cross section indicated by the dotted line in Fig.~\ref{figure2} (a).          
After a nitride layer (green), it is deposited the Poly0 material (red), panel (b).
The first micro-coil is fabricated with a suitable mask by using photolithography 
and plasma etching, panel (c).
Then, the OX1 and OX2 are deposited [panel(d)] and these layers 
are again photolithographically patterned
to define the via connecting the Poly0 layer, panel (e).
To finish off, after depositing the Poly2 layer [panel(f)],
a final photolithography and plasma etching are performed to build the second micro-coil, panel (g).
Note that in this last step, due the conformal deposition of Poly2, the previously fabricated via is filled
and both micro-coils are electrically connected.
In this design we have omitted the Poly1 layer as only two conductive materials separated by an insulator are needed.  
In this way, the two micro-coils are separated by an oxide layer of $2.75$ $\mu$m
(the separation in Fig.~\ref{figure1} is exaggerated to facilitate visualization). 
Finally, gold metal is used to make the two square contacts in Fig.~\ref{figure2} (a).
 
%........................................................................................
\section{Experimental setup \label{setup}}

\begin{figure}[t!]
\begin{center}
\includegraphics[width=5cm,clip=true]{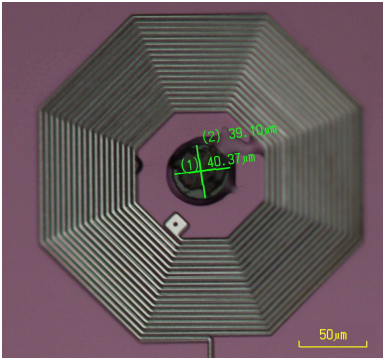}
\caption{Photograph of the micron-sized disk of BSCCO glued on the center of a micro-coil.}
\label{figure3}
\end{center}
\end{figure}

The superconducting sample that we have studied, a micron-sized disk of BSCCO single crystal, 
was fabricated by optical lithography and physical ion milling \cite{Dolz2015}.  
It has a thicknesses of $h=1 \ \mu$m, a diameter of $d=40 \ \mu$m,
and their $ab$ superconducting layers are parallel to the plane of the disk.
The sample was placed on the center of a micro-coil using a hydraulic micromanipulator 
under an optical microscope, and glued to it with a drop 
of Apiezon$\textcopyright$ N grease.  
Figure~\ref{figure3} shows a photograph of this system.

We apply an ac magnetic field normal to the surface of the sample, 
using a circular primary coil of copper with $450$ turns,
external and internal diameters of $16.5$ and $5$ mm, respectively, and height of $6.5$ mm.
The micro-sensor with the sample was placed at $5.5$ mm from the center of this coil,
and the whole system was cooled under vacuum inside a helium closed-cycle cryogenerator.  
Further, the experiment was carried out without applying a dc magnetic field. 
The signal produced by the sensor was measured using a dual lock-in amplifier (Signal Recovery 7280).

In practice, it can be very difficult to use the micro-sensor
as the two micro-coils are not perfectly equal 
(the thicknesses of Poly0 and Poly2 are different).
The electrical resistance of the doped polysilicon layers strongly depend on temperature
and therefore a background masks the signal of interest.  
For example for in a  typical measurement, between $T=100$ K and $T=80$ K,
the voltage measured at the ends of a micro-sensor (without a magnetic sample) changes tens of microvolts, 
while the magnitude of the signal that should be generated when the sample becomes magnetized
would be of the order of a few microvolts. 

\begin{figure}[t!]
\begin{center}
\includegraphics[width=8cm,clip=true]{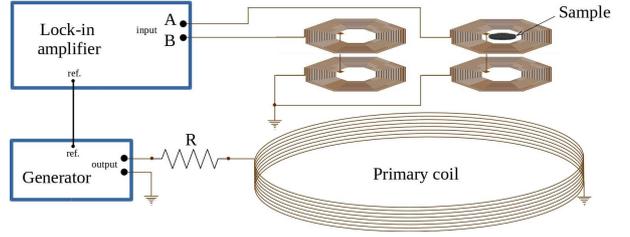}
\caption{Schematic diagram of the experimental setup. 
Two identical micro-sensors, one without and another with a magnetic sample, 
are subject to the same ac magnetic field produced by the primary coil.
This coil has a larger size than each micro-coil and 
is connected (through a limiting resistor of $R=120 \ \Omega$) to a signal generator.
The rms voltage is measured by a lock-in amplifier 
operating in a differential mode
whose reference is provided by the same signal generator.}
\label{figure4}
\end{center}
\end{figure}

To overcome this problem we use the experimental setup shown in Fig.~\ref{figure4}.
Two micro-sensors, one of them empty used as reference 
and the other having the sample under study, are subject to the same 
alternating field produced by the primary coil. 
The electrical connection shown in this figure, 
where the lock-in amplifier is operating in a differential mode (A-B),
allows to attenuate the background signal. 

Finally, the measurements were made between $T=100$ K and $T=30$ K
at a slow cooling rate of $0.1$ K/min, to keep the sample thermally coupled 
to the cryogenerator.   

%............................................................................................
\section{Results and discussion \label{results}}

A simple calculation allows to link the signal measured by the lock-in amplifier, 
with the {\em external} complex magnetic susceptibility \cite{Hein1991,Nikolo1995} of this particular sample.  
According to the Faraday's law, and supposing that the ac magnetic field $H$
produced by the primary coil is uniform and normal to the plane of the micro-sensor,  
the induced electromotive force in the micro-coil without sample (denoted with the subscript ``1'') is  
\begin{equation}
\varepsilon_1=-\frac{d\Phi_1}{dt}=-A_\mathrm{eff} \mu_0 \frac{dH}{dt}. \label{fem1}
\end{equation}
Here, $A_\mathrm{eff}$ is the effective area of the micro-coil (the sum of the areas of $N=18$ octagons)
and $\mu_0$ is the vacuum permeability. 
On the other hand, as the magnetic flux density inside the sample is $B= \mu_0 (H+M)$, 
with $M$ being its magnetization, the flux through the other micro-coil (denoted with the subscript ``2'')
is approximately $\Phi_2=(A_\mathrm{eff}-NA)\mu_0 H + NA \mu_0 (H+M)=A_\mathrm{eff}\mu_0 H + NA \mu_0 M$,
where $A$ is the area of the sample.
Consequently, 
\begin{equation}
\varepsilon_2=-\frac{d\Phi_2}{dt}=-A_\mathrm{eff} \mu_0 \frac{dH}{dt}-NA \mu_0  \frac{dM}{dt}  \label{fem2}
\end{equation}
and the total induced electromotive force in the micro-sensor is
\begin{equation}
\varepsilon =\varepsilon_2-\varepsilon_1=-NA \mu_0  \frac{dM}{dt}. \label{fem}
\end{equation}  
Note that to obtain the last equation [and in particular Eq.~(\ref{fem1})],  
we assume that the sample magnetization does not disturb the flux through the micro-coil number ``1''.
As we will see below, this approximation is quite good since the results obtained by us 
agree very well with the experimental data reported in the literature.

In response to a harmonic magnetic field $H=H_0 \cos(\omega t)$,
with $H_0$ being its amplitude, $\omega$ the angular frequency and $t$ the time,
the magnetization of the sample will show a phase lag $\theta$ behind this applied field.
If $\chi$ is the module of the external complex susceptibility 
(defined through the expression $M=\chi H$), then
\begin{equation} 
M=\chi H_0 \cos(\omega t - \theta)=H_0 [\chi' \cos(\omega t)+\chi'' \sin(\omega t)], \label{mag}
\end{equation}
where $\chi'=\chi \cos(\theta)$ and $\chi''=\chi \sin(\theta)$ are, respectively,
the real and imaginary parts of the external susceptibility.
As it is well known, $\chi'$ is a measure of the magnetic shielding of the sample
while $\chi''$ is related to dissipation processes \cite{Gomory1997}.
Using Eq.~(\ref{mag}), the total induced electromotive force Eq.~(\ref{fem}) can be written as
\begin{equation}
\varepsilon = \varepsilon_0 \chi \sin(\omega t - \theta)=\varepsilon_0 [\chi' \sin(\omega t)-\chi'' \cos(\omega t)], 
\end{equation}
where $\varepsilon_0 =NA \mu_0 H_0 \omega$.
With a dual lock-in amplifier it is possible to measure $v=\varepsilon/\sqrt{2}$, the rms values of this voltage,
but also $v_x$ and $v_y$, their in-phase and quadrature components. 
In our experimental setup (see Fig.~\ref{figure4}) the signal produced by the generator, 
which is proportional to $\cos(\omega t)$, is used as an external reference for the lock-in amplifier.
From this wave the instrument generates its own reference signal at frequency 
$\omega$ with a fixed phase shift of $\phi_r$.
As the electrical connections in both the primary and secondary circuits add a new phase shift of $\phi_c$, 
the (effective) reference signal used to compute $v_x$ and $v_y$ will be $\cos(\omega t + \phi)$,
with $\phi=\phi_r+\phi_c$.
Therefore, it is easy to shows that \cite{Nikolo1995}
\begin{equation}
v_x=-v_0 \chi' \sin(\phi) -v_0 \chi'' \cos(\phi)
\end{equation}
and
\begin{equation}
v_y=-v_0 \chi' \cos(\phi) +v_0 \chi'' \sin(\phi),
\end{equation}
where $v_0=\varepsilon_0/\sqrt{2}$.
Finally, from these equations it follows that 
\begin{equation}
\chi'=-\frac{v_x}{v_0} \sin(\phi)-\frac{v_y}{v_0} \cos(\phi) \label{C1}
\end{equation}
and
\begin{equation}
\chi''=-\frac{v_x}{v_0} \cos(\phi)+\frac{v_y}{v_0} \sin(\phi). \label{C2}
\end{equation}

In order to calculate the real and imaginary parts of $\chi$,
first it is necessary to determine the value of $\phi$.
Here, we use the fact that $\chi''=0$ for a superconducting sample well below the critical temperature, 
i. e., well deep in the Meissner phase no dissipation will be observed. 
Therefore, from Eq.~(\ref{C2}) we determine that
\begin{equation}
\phi=\arctan \frac{v_x^\ast}{v_y^\ast}, \label{phi}
\end{equation}
where $v_x^\ast$ and $v_y^\ast$ are, respectively, the in-phase and quadrature components of $v$ 
measured at the lowest temperature reached in the experiment.   

\begin{figure}[t!]
\begin{center}
\includegraphics[width=7cm,clip=true]{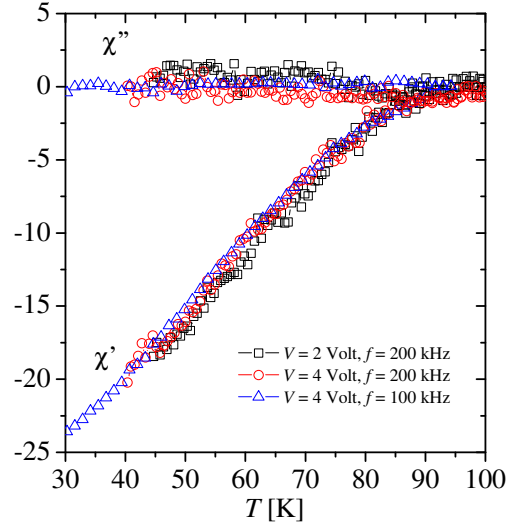}
\caption{The real and imaginary parts of the external susceptibility
as function of temperature, measured for two different voltages and frequencies as indicated.}
\label{figure5}
\end{center}
\end{figure}

Using Eqs.(\ref{C1}-\ref{phi}), we calculate the external susceptibility of the micron-sized disk of BSCCO.  
With the signal generator we apply voltages of amplitude $V$ ranging between 1 and 4 Volts 
(which correspond to ac magnetic fields of amplitude $H_0$ between $0.1$ and $0.4$ Gauss),
and frequencies $f$ of 100 and 200 kHz.
Figure~\ref{figure5} shows the real and imaginary parts of $\chi$ obtained in our experiments
for $V=2$ Volt and $V=4$ Volt at $f=200$ kHz, and for $V=4$ Volt at $f=100$ kHz.
Although the measurements are significantly affected by noise,
the curves agree very well with each other for this set of parameters
and clearly show, as expected, the transition to the Meissner phase. 
$\chi'$ becomes negative around 90 K, the critical temperature in zero dc field 
previously determined in experiments performed with micro-Hall sensors \cite{Dolz2015,Dolz2014}.
Nevertheless, the typical dissipation peak in $\chi''$ is not clearly observed,
because the magnitude of this signal should be at most of $0.025$ \cite{Arribere1993},
a value much smaller than the noise levels we recorded. 
        
In order to check the validity of our measurements, 
we compare with experimental data reported in the literature.
The two parts of the {\em internal} complex magnetic susceptibility, $\chi'_\textrm{int}$ and $\chi''_\textrm{int}$,
of samples of BSCCO single crystal have been measured
for zero dc field \cite{Arribere1993,Pastoriza1992}.  
The module of this quantity, which is a characteristic of a given material, 
can be defined through the expression $M=\chi_\textrm{int} H_\textrm{int}$,
where $H_\textrm{int}=H-DM$ is the internal field 
and $D$ the demagnetizing factor of the sample along the direction of the applied ac field. 
Neglecting the imaginary components of both, the external and internal susceptibilities
(since for zero dc field $\chi'' \ll \chi$ and $\chi''_\textrm{int} \ll \chi_\textrm{int}$), 
their real parts are related by the equations \cite{Hein1991}  
\begin{equation}
\chi'= \frac{\chi'_\textrm{int}}{\Big( 1+D\chi'_\textrm{int} \Big)} \label{relation1} 
\end{equation}
and
\begin{equation}
\chi_\textrm{int}= \frac{\chi}{\Big( 1-D\chi \Big)}.   \label{relation2} 
\end{equation}
In addition, since our sample has a large aspect ratio of $m=d/l=40$, 
its demagnetizing factor along the perpendicular direction 
to the disk plane is $D \approx 0.961$ \cite{Cullity}.

\begin{figure}[t!]
\begin{center}
\includegraphics[width=7cm,clip=true]{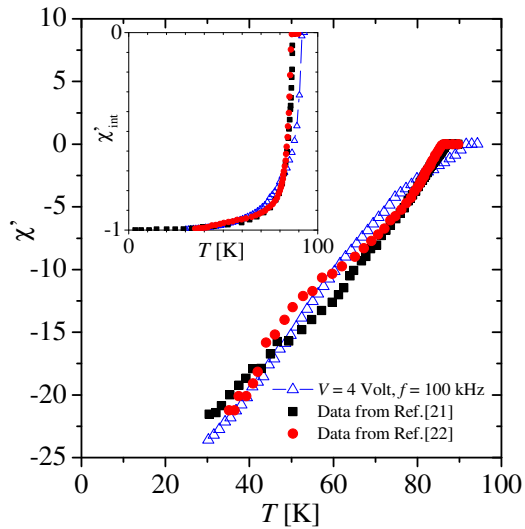}
\caption{Comparison between the real part of the external susceptibility 
measured in our experiments for $V=4$ Volt at $f=100$ kHz,
and those calculated from data of Refs.\cite{Arribere1993,Pastoriza1992} as indicated.
Inset shows the real part of the internal susceptibility for these same measurements.}
\label{figure6}
\end{center}
\end{figure}

In Fig.~\ref{figure6} we show the real part of the external 
susceptibility measured in our experiments for $V=4$ Volt at $f=100$ kHz,
and those calculated from data of Refs.\cite{Arribere1993,Pastoriza1992} 
using the Eq.~(\ref{relation1}). 
Even though these curves have been obtained by different techniques,
they agree very well with each other showing that our micro-sensor 
is able to detect the superconducting transition of micron-sized samples.
In the inset we show an equivalent comparison for the real part of the internal susceptibility 
[form our data for $\chi'$, we calculate the corresponding $\chi'_\textrm{int}$ using Eq.~(\ref{relation2})].
We see again a good agreement despite the differences
between the critical temperatures of each sample.

%............................................................................................
\section{Conclusions \label{conclusions}}

We have implemented an induction micro-sensor which can detect 
the magnetic-flux expulsion of mesoscopic superconductor samples.  
This device consists of two octagonal planar parallel micro-coils 
and was built using the multiuser commercial process PolyMUMPs of Memscap Company.
We study a micron-sized disk of BSCCO single crystal which was fabricated by optical lithography and physical ion milling.
We measure the real and imaginary components of the external complex susceptibility of this sample.
The Meissner transition is detected at a critical temperature of $T_c \approx 90$K in zero dc field,
in agreement with previous experiments performed with micro-Hall sensors \cite{Dolz2015,Dolz2014}. 
Unfortunately, the noise levels that we recorded do not allow to distinguish the typical dissipation peak in $\chi''$.
Finally, comparing our measurements with experimental data reported in the literature,
the good performance of our micro-sensor was verified. 
 
\section*{Acknowledgments}
This work was supported in part by CONICET under project number PIP 112-201301-00049-CO, 
by FONCyT under project PICT-2013-0214, and by Universidad Nacional de San Luis under 
project PROICO P-31216 (Argentina).

%............................................................................................
\end{document}